# The Physics Show - A community college delivers BIG on community outreach

**Frank Cascarano and David Marasco**

Founded in 2007, the Foothill College Physics Show has served nearly a quarter of a million attendees in the two decades that have followed. This demo show features both performances for the public and field trips for students from local Title 1 schools. The college's students play an important role, acting as both on-stage talent, leading tours of the college, and helping build equipment. From a small beginning, it now hosts over twenty-five thousand attendees a year, and is an important part of the college's outreach efforts.

Originally inspired by videos of Clint Sprott's Wonders of Physics program[1] at Wisconsin-Madison, and later by The Physics Force at University of Minnesota,[2] in 2007 the Foothill College Physics Department's new faculty wished to provide a similar experience for the local community, inspiring the youth with physics. In spring of that year a free show took place in the college's largest lecture hall before 175 people, mainly families. Included in the act were classic demonstrations such as liquid nitrogen and a Ruben's Tube with the bed of nails as a finale. While topics are rotated in and out of the set list on a three-year cycle, this initial performance established the template. The event is targeted at elementary and middle school-aged children, lasts for roughly 75 minutes, and closes with the bed of nails.

The following year consisted of four performances, which "sold out" by word of mouth almost instantly. The third year saw a move to the campus's main theater, which has a capacity of nearly a thousand seats. This marked a transition from a small, informal production to a large-scale effort that employs professional theater technicians, and one that uses live video and a projection screen.  Since then the number of performances has grown to the point where The Physics Show is now the theater's anchor tenant, delivering more attendees than any of the college's other attractions.  Around this time there was another shift, moving to a paid-admission model. Prior to that a limited number of tickets were released online on a first-come-first served basis. As tickets were free, people who reserved seats could no-show at no cost. With a thousand-person theater, even a modest no-show rate entails many empty seats. Charging for admission both allowed the program to cover theater labor costs and increase actual attendance.

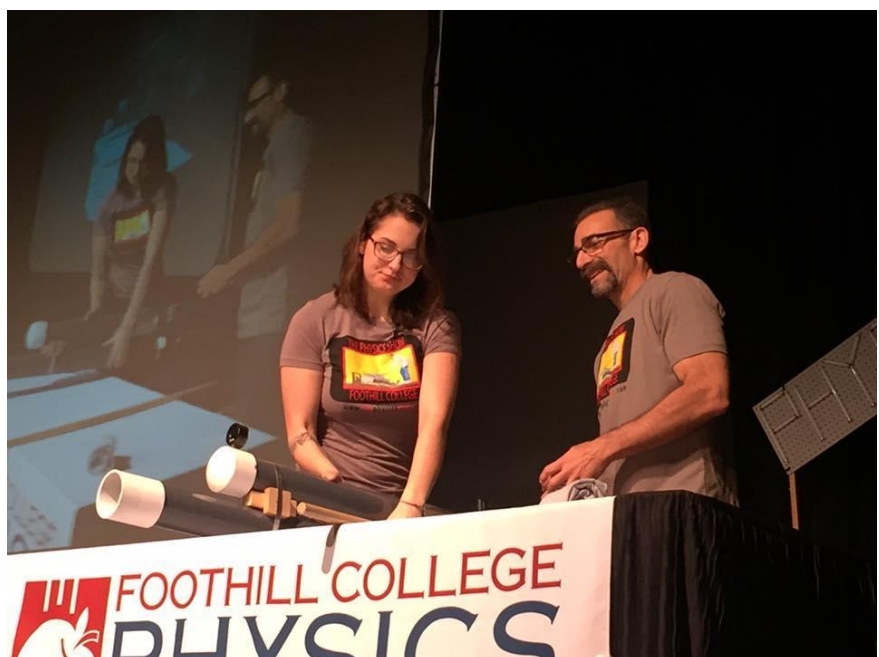
*Figure 1 - Student and instructor presenting onstage*

**Field Trip Shows - Outreach to Title 1 Schools**
For the first several years there were performances on a weekend or two a year. After one of these a person remarked "...your demos are great, I haven't thought about these things since grad school." This comment triggered an introspection of our goals. It was recognized that due to a Silicon Valley location, many of the parents bringing children to the show worked at Google, Facebook, Stanford, Apple, and other tech-heavy environments. These children would be steered to both science and college, whether or not they attended a Physics Show. The decision was made to raise ticket prices, and to use the proceeds to do outreach to populations that did not have the privileges enjoyed by the then-current attendees.

The Physics Show invites classes from local Title 1 schools, and has slowly grown to hosting five thousand students annually on field trips to the college. The program supports the cost of the buses, performs the show, provides a tour of the college's campus (lead by a current Foothill College student), and gives each child a Foothill College Physics Show shirt. Teachers often share that this is the only field trip the school has for the year, as the schools lack budget for field trips. For many kids this is the first time they step onto a college campus; they see that college is an inviting place that could be part of their future. With every child in their class getting a shirt, on any given school day the odds are that they will see somebody wearing a Foothill College Physics Show shirt, a constant reminder of both the possibility of college and the excitement of physics. Over forty thousand underserved youth have been hosted by the field trip program, and which has become an important player in the college's outreach efforts. Roughly fifty Foothill College students a year are employed as college tour guides for the field trip shows.

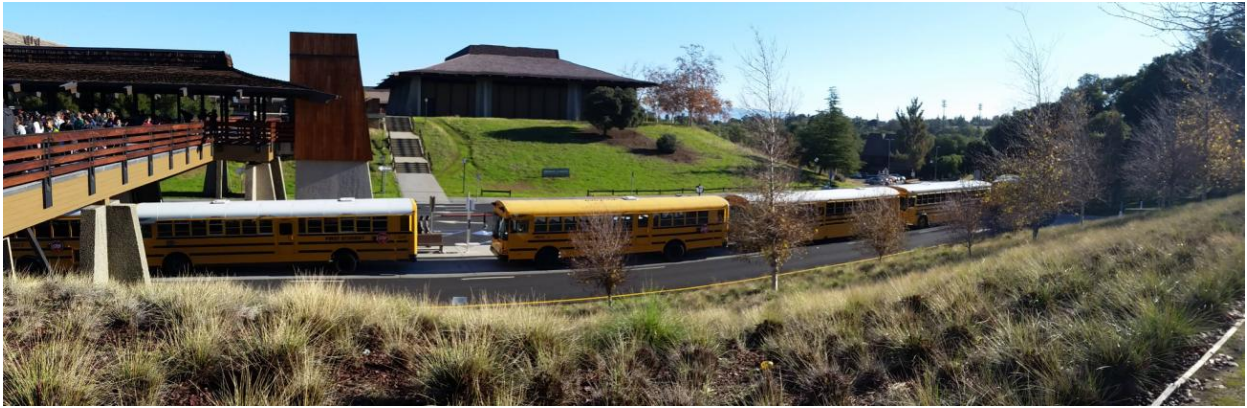
*Figure 2 - School buses and children arriving for a field trip show*

**Student Involvement**

Student participation is a critical component, with student presenters on stage explaining the physics in many of the acts. This involves the additional tasks of training and rehearsing with the student presenters, but importantly puts people on the stage who are more relatable to the young attendees. This reinforces the idea that they can do physics when they get older. In fact, many recent presenters saw the Physics Show when they were younger (this also holds true for students in our classes). Over fifty students have served as on-stage performers, and most have gone on to have solid careers in STEM. Another audience is parents, who get to see confident and capable community college students. As this represents a major time commitment for students, all of our community college students are paid from the Physics Show budget (faculty volunteer their time and the college donates the use of the facility).

While some of the demo equipment is purchased from vendors, much of it is built with the assistance of the Science and Engineering Club. Some parts of the act are enabled by a close-up camera, but in a thousand-person theater it is important to "super-size" many of the demos. With help from community volunteers, the students have made a wide variety of large projects, from magnetic credit card readers to a giant pendulum snake. This work provides valuable planning, crafting, and troubleshooting skills to our students, and also provides an important role for those who feel less comfortable about acting on a stage. A student who later became a performer also called out the camaraderie, "I got involved with the Engineering Club because it seemed fun and I thought it would give me practical skills. Through that, I got involved with the Physics Show. It's a really good community of students that come together."[3]

Students bring a fresh set of eyes to both the material and our communication. As an example, this year one of the presenters suggested that a professor wear a ballerina's tutu during a standard conservation of angular momentum demo, both reinforcing the practical application and adding to the show's playful atmosphere. This led to a discussion with a non-binary identifying crew member, to ensure that the framing was inclusive rather than harmful. The result, "Dr. Marasco wants you to know that anyone can wear a tutu and that anyone can do physics," was a line that many children remembered.

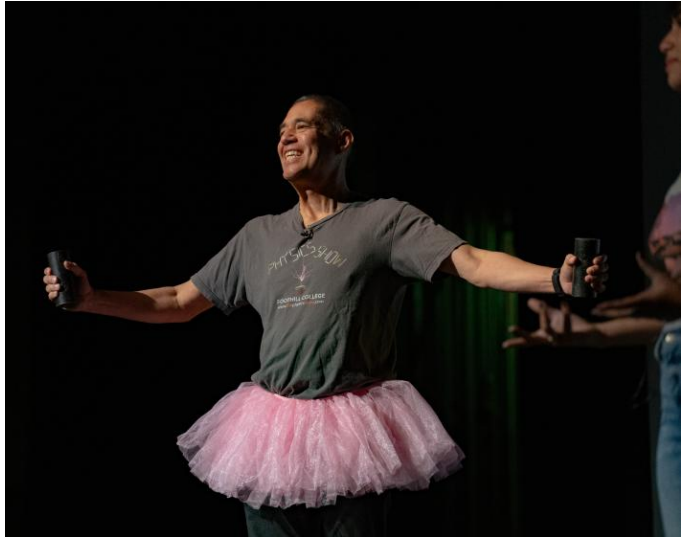
*Figure 3 - Conserving angular momentum in a tutu.*

**Conclusion**

Since its founding, the Foothill College Physics Show has developed from 175 fans in a lecture hall to a powerful outreach program with an annual attendance of over 25,000. After losing a year to COVID-19, it has successfully rebooted and ramped up beyond pre-pandemic levels. It holds roughly thirty performances a year with both shows for the public and field trip days. The show greatly benefits from the contributions of physics students, both on stage and behind the scenes.

**References**

1. C. Sprott, The Wonders of Physics (U. Wisconsin, Madison, WI), http://sprott.physics.wisc.edu/wop.htm.

2. Physics Force (U. Minnesota Twin Cities, Minneapolis, MI), https://physicsforce.umn.edu/

3. Sarwari, Khalida "Foothill's Physics Show puts the sizzle in science," San Jose Mercury News, November 23, 2016, https://www.mercurynews.com/2016/11/23/foothills-physics-show-puts-the-sizzle-in-science/